\documentclass[twocolumn,showpacs,amsmath,amssymb,10pt]{revtex4}
\usepackage{amsfonts}
\usepackage{bm}

\newcommand{\ot}{{\,\otimes\,}}
\newcommand{{\Cd}}{{\mathbb{C}^d}}

\def\oper{{\mathchoice{\rm 1\mskip-4mu l}{\rm 1\mskip-4mu l}%
{\rm 1\mskip-4.5mu l}{\rm 1\mskip-5mu l}}}
\def\<{\langle}
\def\>{\rangle}

\begin{document}
\title{\textbf{Quantum states with strong positive partial transpose}}
\author{Dariusz Chru\'sci\'nski, Jacek Jurkowski  and Andrzej Kossakowski\thanks{email:
darch@phys.uni.torun.pl} }
\affiliation{Institute of Physics, Nicolaus Copernicus University,\\
Grudzi\c{a}dzka 5/7, 87--100 Toru\'n, Poland}

\begin{abstract}

We construct a large class of bipartite $M \ot N$ quantum states
which defines a proper subset of states with positive partial
transposes (PPT).  Any state from this class is PPT but the
positivity of its partial transposition is recognized  with
respect to canonical factorization of the original density
operator.  We propose to call elements from this class states with
strong positive partial transposes (SPPT). We conjecture that all
SPPT states are separable.
\end{abstract}

\maketitle

Quantum entanglement is one of the most remarkable features of
quantum mechanics and it leads to powerful applications like
quantum cryptography, dense coding and quantum computing
\cite{QIT,Horodecki-review}.

One of the central problems in the theory of quantum entanglement
is to check whether a given density matrix describing a quantum
state of the composite system is separable or entangled. Let us
recall that a state represented by a density operator $\rho$
living in the Hilbert space $\mathcal{H}_A \ot \mathcal{H}_B$ is
separable iff $\rho$ is a convex combination of product states,
that is, $\rho = \sum_k p_k \rho^{(A)}_k \ot \rho^{(B)}_k$, with
$\{p_k\}$ being a probability distribution, and $\rho^{(A)}_k, \
\rho^{(B)}_k$ are density operators describing states of subsystem
$A$ and $B$, respectively \cite{Werner}.

 There are several operational criteria which enable
one to detect quantum entanglement (see e.g.
\cite{Horodecki-review} for the recent review). The most famous
Peres-Horodecki criterion \cite{Peres,PPT} is based on the partial
transposition: if a state $\rho$ is separable then its partial
transposition $\rho^{{\rm T}_A} = ({\rm T} \ot \oper)\rho$ is
positive (such states are called PPT state). The structure of this
set is of primary importance in quantum information theory.
Unfortunately, this structure is still unknown, that is, one may
easily check whether a given state is PPT but we do not know how
to construct a general quantum state with PPT property.

Recently \cite{III,CIRCULANT} we proposed  large classes of states
where the PPT property is very easy to check. In the present
Letter we propose a new class of states which are PPT by the very
construction. This construction is based on the block structure of
any density matrix living in the tensor product $\mathcal{H}_A \ot
\mathcal{H}_B$, that is, a density matrix in $\mathbb{C}^M \ot
\mathbb{C}^N$ may be considered as  $M \times M$ matrix with $N
\times N$ blocks. Partial transposition is an operation which acts
on blocks and we show how to organize blocks to have a density
matrix with PPT property. We propose to call PPT states
constructed this way a strong PPT states (SPPT). Interestingly,
known examples of SPPT states turn out to be separable. This
observation supported by some numerical investigations encouraged
us to conjecture that all SPPT states are separable.

The Letter is organized as follows: for pedagogical reason we
start with $M=2$ and arbitrary (but finite) $N$. This construction
easily generalizes for arbitrary $M > 2$. We finish with some
conclusions.

{\it 1. $2 \ot N$ systems}. Such systems are of primary importance
in quantum information theory and they were extensively analyzed
in \cite{2N}. It is clear that an (unnormalized) state of a
bipartite system living in $\mathbb{C}^2 \ot \mathbb{C}^N$ may be
considered as a block $2 \times 2$ matrix with $N \times N$
blocks. Positivity of $\rho$ implies that $\rho =
\mathbf{X}^\dagger \mathbf{X}$ for some $2N \times 2N$ matrix $\bf
X$. Again, this matrix may be considered as a block $2\times 2$
matrix with $N\times N$ blocks. Consider now the following class
of upper triangular block matrices $\bf X$:
\begin{equation}\label{X}
\mathbf{X} = \left( \begin{array}{c|c} X_1 & SX_1  \\ \hline
  0 & X_2  \end{array} \right)\ ,
\end{equation}
with arbitrary $N \times N$ matrices $X_1,X_2$ and $S$.  One finds
\begin{equation}\label{XX-2}
    \rho = \mathbf{X}^\dagger \mathbf{X} =  \left( \begin{array}{c|c} X_1^\dagger  X_1 & X_1^\dagger  S X_1  \\
\hline  X_1^\dagger  S^\dagger  X_1 & X_1^\dagger  S^\dagger  S
X_1 + X_2^\dagger  X_2
\end{array} \right) \ ,
\end{equation}
and  its partial transposition is given by
\begin{equation}\label{T-rho}
    \rho^{{\rm T}_A} =
\left( \begin{array}{c|c} X_1^\dagger  X_1 & X_1^\dagger  S^\dagger  X_1  \\
\hline  X_1^\dagger S X_1 & X_1^\dagger  S^\dagger  S X_1 +
X_2^\dagger X_2
\end{array} \right) \ .
\end{equation}
Clearly, $\rho$ is PPT iff there exists $\bf Y$ such that
$\rho^{T_A} = {\bf Y}^\dagger {\bf Y}$. The choice of $\bf Y$ (if
it exists) is highly nonunique.  Note, however, that there is a
`canonical' candidate for $2N \times 2N$ matrix  $\bf Y$ defined
by (\ref{X}) with $S$ replaced by $S^\dagger$, that is
\begin{equation}\label{Y}
{\bf Y} = \left( \begin{array}{c|c} X_1 & S^\dagger X_1  \\
\hline  0 & X_2  \end{array} \right) \ ,
\end{equation}
and hence
\begin{equation}\label{YY}
    {\bf Y}^\dagger {\bf Y} =
\left( \begin{array}{c|c} X_1^\dagger  X_1 & X_1^\dagger  S^\dagger  X_1  \\
\hline  X_1^\dagger   S X_1 & X_1^\dagger  S S^\dagger  X_1 +
X_2^\dagger X_2
\end{array} \right) \ .
\end{equation}
Now, we say that a state $\rho = {\bf X}^\dagger {\bf X}$ with
$\bf X$ defined in (\ref{X})  has {\bf strong positive partial
transpose} (SPPT) iff $\rho^{T_A} = {\bf Y}^\dagger {\bf Y}$ with
$\bf Y$ defined in (\ref{Y}).

It is therefore clear that a $2 \ot N$ state $\rho$ is SPPT if and
only if
\begin{equation}\label{SS}
 X_1^\dagger S^\dagger  S X_1 = X_1^\dagger S S^\dagger X_1 \ .
\end{equation}
Note, that if $S$ is normal, i.e. $S^\dagger  S =  S S^\dagger$,
then $\rho$ is necessarily SPPT. It was proved in \cite{2N} that
if the rank of $\rho$ is $N$, then PPT implies separability. Now,
any PPT $\rho$ of rank   $N$ may be constructed via (\ref{XX-2})
with $X_1 = \mathbb{I}$, $X_2=0$ and a normal matrix $S$ giving
rise to (so called canonical $2\ot N$ form \cite{2N})
\[  \rho =  \left( \begin{array}{c|c} \mathbb{I} &   S  \\
\hline   S^\dagger  & S^\dagger  S
\end{array} \right) \ . \]
Due to normality of $S$ it does belong to our class, i.e. any rank
$N$ PPT state in $2\ot N$ is both SPPT and separable. Another
example of SPPT states is provided by hermitian (and hence normal)
$S$. It implies $\rho^{T_A}=\rho$. It is well known \cite{2N} that
for $2\ot N$ systems this condition is sufficient for
separability. Hence, for $2 \ot N$ case all states defined by
arbitrary $X_1\, ,X_2$ and arbitrary but hermitian $S$ are SPPT
from (\ref{SS}) and separable due to \cite{2N}.

Consider other well known examples in $2 \ot N$. The celebrated
Werner state \cite{Werner} in $2 \ot 2$ is SPPT if and only if it
is maximally mixed, i.e. $ \frac 14\, \mathbb{I} \ot \mathbb{I}$.
The same is true for the isotropic state in $2 \ot 2$. The seminal
Horodecki entangled PPT state \cite{PPT} in $2 \ot 4$
parameterized by $b\in [0,1]$ belongs to our class iff $b=0$ (for
$b=0,1$ Horodecki state is separable). In a recent paper
\cite{CIRCULANT} we constructed a class of so called circulant
states in $N \ot N$. For $N=2$ they are given by
\begin{equation}\label{2C}
    \rho = \left( \begin{array}{cc|cc}
    a_{11} & 0 & 0 & a_{12} \\
    0      & b_{11} & b_{12} & 0 \\ \hline
    0      & b_{21} & b_{22} & 0 \\
    a_{21} & 0 & 0 & a_{22} \end{array} \right)\ ,
\end{equation}
where $[a_{ij}]$ and $[b_{ij}]$ are $2 \times 2$ positive
matrices. Partially transposed $\rho$ has the same structure but
with $[a_{ij}]$ and $[b_{ij}]$ replaced by $[\widetilde{a}_{ij}]$
and $[\widetilde{b}_{ij}]$
\[  \widetilde{a} = \left( \begin{array}{cc}
    a_{11} & b_{21} \\
    b_{12} & a_{22} \end{array} \right) \ , \ \ \ \
\widetilde{b} = \left( \begin{array}{cc}
    b_{11} & a_{21} \\
    a_{12} & b_{22} \end{array} \right)\ .  \]
Now, $\rho$ is PPT iff $\widetilde{a}\geq 0$ and $\widetilde{b}
\geq 0$. It is not difficult to see that a circulant $2 \ot 2$ PPT
state is SPPT iff $|a_{12}|=|b_{12}|$. A nice example of circulant
state is provided by orthogonally invariant state \cite{Werner2},
that is, a 2-qubit state $\rho$ satisfying $U \ot U \rho = \rho U
\ot U$, with $U\in U(2)$ and $\overline{U}=U$:
\begin{equation}\label{2C-T}
\rho = \frac 14 \left( \begin{array}{cc|cc}
    a +2b & \cdot & \cdot & 2b-a \\
    \cdot      & a+2c & a-2c & \cdot \\ \hline
    \cdot      & a-2c & a+2c & \cdot \\
    2b-a & \cdot & \cdot & a+2b \end{array} \right)\ ,
\end{equation}
where $a,b,c\geq 0$ and $a+b+c=1$. It is easy to see that  $\rho$
is PPT iff  $ b,c \leq 1/2$ \cite{Werner2}. Moreover, $\rho$ is
SPPT iff it is PPT and $b=c$. Hence SPPT states define a
1-parameter family within 2-parameter class of PPT states.

{\it 2. General $M \ot N$ systems}.   The above construction may
be easily generalized for an arbitrary
 bipartite system living in $\mathbb{C}^M \ot \mathbb{C}^N$. Now, a state $\rho$
 may be considered as an $M \times M$ matrix with entries being $N \times N$ matrices.
  Positivity of $\rho$ implies that $\rho = {\bf X}^\dagger {\bf X}$ for some $MN \times MN$
matrix $\bf X$ --- a block $M\times M$ matrix with $N\times N$
blocks. Let us consider the following class of upper triangular
block matrices $\bf X$: diagonal blocks $X_{ii} = X_i$ and $X_{ij}
= S_{ij}X_i$ for $i<j$
\begin{equation*}\label{X-3}
{\bf X} = \left( \begin{array}{c|c|c|c|c} X_1 & S_{12}X_1 & S_{13}X_1 & \ldots & S_{1M}X_1 \\
\hline  0 & X_2 & S_{23}X_2 &\ldots & S_{2M}X_2 \\ \hline \vdots &
\vdots & \ddots & \vdots & \vdots \\ \hline 0 & 0 & 0 & X_{M-1} &
S_{M-1,M}X_{M-1} \\ \hline 0 & 0 & 0 & 0 & X_M
\end{array} \right) \ ,
\end{equation*}
where $X_k$ and $S_{ij}\ (i<j)$ are $N \times N$ matrices. Simple
calculation gives for diagonal blocks
\begin{eqnarray}\label{Xa}
\rho_{11} &=& X_1^\dagger X_1 \ , \nonumber \\
\rho_{22} &=& X_1^\dagger S_{12}^\dagger S_{12} X_1 + X_2^\dagger X_2 \ , \nonumber \\
\rho_{33} &=& X_1^\dagger S_{13}^\dagger S_{13} X_1 + X_2^\dagger
S_{23}^\dagger S_{23} X_2 +
 X_3^\dagger X_3 \ , \nonumber \\
  &\vdots&   \\
\rho_{MM} &=& \sum_{k=1}^{M-1} X_k^\dagger S_{kM}^\dagger S_{kM}
X_k +  X_M^\dagger X_M \ ,\nonumber
\end{eqnarray}
Off-diagonal blocks are defined as follows: for $i=1$
\begin{equation}\label{Xb}
    \rho_{1j} =  X_1^\dagger  S_{1j} X_1 \ ,
\end{equation}
and for $1 < i < j$
\begin{equation}\label{Xc}
    \rho_{ij} = \sum_{k=1}^{i-1} X_k^\dagger  S^\dagger_{ki} S_{kj} X_k + X_i^\dagger  S_{ij} X_i \
    .
\end{equation}
 Partially transposed $\rho^{T_A}$ is therefore
given by the following block matrix: diagonal blocks
\begin{equation}\label{}
   \rho^{T_A}_{ii} = \rho_{ii}\ ,
\end{equation}
and off-diagonal blocks: for $i=1$
\begin{equation}\label{}
    \rho^{T_A}_{1j} = \rho^\dagger_{j1} =  X_1^\dagger  S^\dagger_{1j} X_1 \ ,
\end{equation}
and for $1 < i < j$
\begin{equation}\label{}
    \rho^{T_A}_{ij} = \rho^\dagger_{ji} = \sum_{k=1}^{i-1} X_k^\dagger  S^\dagger_{kj} S_{ki} X_k + X_i^\dagger  S^\dagger_{ij} X_i \
    .
\end{equation}
Now, in analogy to $2 \ot N$ case we say that  $\rho$ is SPPT iff
 $\rho^{T_A} = {\bf Y}^\dagger {\bf
Y}$ where ${\bf Y}$ is given by the following `canonical' block
matrix
\begin{equation*}\label{X-3}
{\bf Y} = \left( \begin{array}{c|c|c|c|c} X_1 & S^\dagger_{12}X_1 & S^\dagger_{13}X_1 & \ldots & S^\dagger_{1M}X_1 \\
\hline  0 & X_2 & S^\dagger_{23}X_2 &\ldots & S^\dagger_{2M}X_2 \\
\hline \vdots & \vdots & \ddots & \vdots & \vdots \\ \hline 0 & 0
& 0 & X_{M-1} & S^\dagger_{M-1,M}X_{M-1} \\ \hline 0 & 0 & 0 & 0 &
X_M
\end{array} \right) \ .
\end{equation*}
It is clear that blocks $({\bf Y}^\dagger {\bf Y})_{ij}$ are
defined by the same formulae as $({\bf X}^\dagger {\bf X})_{ij}$
with $S_{ij}$ replaced by $S_{ij}^\dagger$
--- formulae (\ref{Xa})--(\ref{Xc}). Therefore, the SPPT condition
$\rho^{T_A} = {\bf Y}^\dagger {\bf Y}$ is equivalent to:

\begin{itemize}

\item  for $j=2,\ldots,M$

\begin{equation}\label{}
\sum_{k=1}^{j-1} X_k^\dagger S_{kj}^\dagger S_{kj} X_k =
\sum_{k=1}^{j-1} X_k^\dagger S_{kj} S^\dagger_{kj} X_k\ ,
\end{equation}

\item for $2\leq i<j=3,\ldots,M$

\begin{equation}\label{}
    \sum_{k=1}^{i-1} X_k^\dagger  S^\dagger_{kj} S_{ki} X_k
     =  \sum_{k=1}^{i-1} X_k^\dagger  S_{ki}  S^\dagger_{kj} X_k \     .
\end{equation}

\end{itemize}

In particular the above conditions are satisfied if
\begin{equation}\label{SS-SS}
S_{ki} S^\dagger _{kj} = S^\dagger _{kj} S_{ki}\ ,
\end{equation}
for $k < i \leq j $. Formula (\ref{SS-SS}) shows that there are
$\frac 12 M(M-1)$ normal matrices $S_{ij}\ (i<j)$ such that each
matrix $S_{ki}$ commutes with $S^\dagger_{kj}$ for $i<k$. It
introduces $\frac 16(M-1)M(M+1)$ independent conditions for
matrices $S_{ij}$. For $M=2$ it reduces to exactly one condition
(\ref{SS}) for one matrix $S$. The special class of SPPT  states
corresponds to a family of hermitian (and hence normal) matrices
$S_{ij}$ satisfying
\[    [S_{ki}\, ,\,  S_{kj}] = 0\ ,\ \ \ \ k < i \leq j  \ . \]
In this case one simply has $ \rho^{T_A} = \rho\,$.

Let us analyze known examples of $M \ot N$ states belonging to our
class of SPPT states. Now, the situation is much more complicated
since our knowledge about general $M \ot N$ case is very limited.

{\it Example 1)} Similarly as in $2 \ot 2$ case both Werner and
isotropic states in $N \ot N$ are SPPT iff they are maximally
mixed.

{\it Example 2)} The seminal Horodecki $3 \ot 3$ PPT but entangled
state \cite{PPT} is SPPT if and only if $a=0$ (in this case it is
separable).

{\it Example 3)} In \cite{III} we have proposed a class of $N \ot
N$ states defined as follows

\begin{equation}\label{}
    {\rho} = \sum_{i,j=1}^N a_{ij} \, |ii\>\<jj| + \sum_{i\neq j
    =1}^N b_{ij}\, |ij\>\<ij|\ ,
\end{equation}
where $[a_{ij}]$ is $N \times N$ positive matrix and $b_{ij}\,
(i\neq  j)$ are positive coefficients. It was shown \cite{III}
that  $\rho$ is PPT iff $|a_{ij}a_{ji}|\leq b_{ij}^2$ for $i\neq
j$.  It turns out that this class contains many well known PPT
    states (for example  an isotropic state are
    there). If $N=3$ this state has the following block form (to have more
transparent picture we represent zeros by dots)
\begin{equation*}\label{3C}
 \hspace*{-.1cm}
  \rho = \left( \begin{array}{ccc|ccc|ccc}
    a_{11} & \cdot & \cdot & \cdot & a_{12} & \cdot & \cdot & \cdot & a_{13} \\
    \cdot& b_{12} & \cdot & \cdot & \cdot& \cdot & \cdot & \cdot & \cdot  \\
    \cdot& \cdot& b_{13} & \cdot & \cdot & \cdot & \cdot & \cdot &\cdot   \\ \hline
    \cdot & \cdot & \cdot & b_{21} & \cdot & \cdot & \cdot & \cdot & \cdot \\
    a_{21} & \cdot & \cdot & \cdot & a_{22} & \cdot & \cdot & \cdot & a_{23}  \\
    \cdot& \cdot & \cdot & \cdot & \cdot & b_{23} & \cdot & \cdot & \cdot  \\ \hline
    \cdot & \cdot & \cdot & \cdot& \cdot &  \cdot&  b_{31} & \cdot & \cdot \\
    \cdot& \cdot & \cdot & \cdot & \cdot& \cdot &  & b_{32} & \cdot  \\
    a_{31} & \cdot& \cdot & \cdot & a_{32} & \cdot& \cdot & \cdot & a_{33}
     \end{array} \right)\ .
\end{equation*}
It is clear that $\rho$ is SPPT iff $a_{ij}=0$ for $i\neq j$, that
is, $\rho$ is diagonal and hence separable. We stress that both
Werner and isotropic states do belong to this class.

{\it Example 4)} In a recent paper \cite{CIRCULANT} we proposed a
class of so called circulant PPT states in $N \ot N$. It is easy
to show that for odd $N$ circulant PPT states are SPPT if and only
if they are diagonal (hence separable). However, for even $N$ we
may have circulant states with more complicated structure (cf.
\cite{CIRCULANT}). Circulant SPPT state for $N=2$ was already
presented in (\ref{2C}). It is not difficult to show that again
SPPT property implies separability.

{\it Conclusions.} We constructed a large class of PPT states in
$\mathbb{C}^M \ot \mathbb{C}^N$ --- we called them SPPT states
since they satisfy one extra condition which is {\it strong}
enough to guarantee PPT.  All known to us examples of such states
turn out to be separable. Moreover, we have strong numerical
evidence (realignment criterion) that SPPT states in $\mathbb{C}^3
\ot \mathbb{C}^3$ are separable. Therefore, we are encouraged to
conjecture that all SPPT states are separable. If this conjecture
is true it gives rise to new sufficient criterion for
separability: if $\rho$ is SPPT, then it is separable.

Note, that constructed states give rise to new family of quantum
channels $\Phi : M_M(\mathbb{C}) \longrightarrow M_N(\mathbb{C})$,
where $M_K(\mathbb{C})$ denotes a set of $K \times K$ complex
matrices. If $e_{ij} = |i\>\<j|$ denotes a base in
$M_M(\mathbb{C})$, then the action of the channel corresponding to
state $\rho$ is given by
\begin{equation}\label{Phi}
    \Phi(e_{ij}) = \rho_{ij}\ ,
\end{equation}
where $\rho_{ij}$ defined in ({\ref{Xa})--(\ref{Xc}) are elements
from $M_N(\mathbb{C})$. Now, if our conjecture about SPPT states
is true any quantum channel  defined via (\ref{Phi}) corresponding
to SPPT state $\rho$ is entanglement breaking \cite{EB1,EB2} (see
also \cite{PEB} for classification of channels), i.e. $(\oper_M
\ot \Phi)P^+_M$ is separable, where $P^+_M$ denotes a projector
onto maximally entangled state in $\mathbb{C}^M \ot \mathbb{C}^M$.
Therefore, as a byproduct we derive a large class of entanglement
breaking quantum channels.

In a recent paper \cite{Kus} authors developed new necessary and
sufficient criterion for separability which is based on the
existence of a set of normal commuting matrices. SPPT states may
therefore provide a laboratory of states where the methods of
\cite{Kus} may be applied. They  may shed new light on the
intricate structure of quantum states of composed systems.

\vspace{.5cm}

{\it Acknowledgments}. This work was partially supported by the
Polish State Committee for Scientific Research.

\end{document}